\documentclass[10pt,preprintnumbers,showpacs,amsmath,amssymb,floatfix,twocolumn]{revtex4}
\usepackage{graphicx}
\usepackage{dcolumn}
\usepackage{bm}
\usepackage{longtable}
\usepackage{graphics}
\usepackage{amssymb}
\usepackage{amsmath}
\usepackage{xspace}
\usepackage{epsfig}
\newcommand {\etal}{\begin{itshape}et al\end{itshape}. }
\begin{document}
\title{A First Principles Density-Functional Calculation of the
Electronic and Vibrational Structure of the Key
Melanin Monomers
\footnote{This work relates to Department of the Navy
Grants N00014-03-1-4115, N00014-02-1-1046 and N00014-03-1-4116
issued by the Office of Naval Research International field office.
The United States Government has a royalty-free license throughout
the world in all copyrightable material contained herein.}}
\author{B. J. Powell,$^{1,}$\footnote{Electronic address: powell@physics.uq.edu.au}
T. Baruah,$^{2,3}$ N. Bernstein,$^{2}$ K. Brake,$^1$ Ross H.
McKenzie,$^1$ P. Meredith$^1$ and M. R. Pederson$^2$}
\affiliation{$^1$Department of Physics, University of Queensland,
Brisbane, Queensland 4072, Australia} \affiliation{$^2$Center for
Computational Materials Science, U.S. Naval Research Laboratory,
Washington, D.C. 20375, USA} \affiliation{$^3$Department of
Physics, Georgetown University, Washington, D.C. 20057, USA}


\begin{abstract}
We report first principles density functional calculations for
hydroquinone (HQ), indolequinone (IQ) and semiquinone (SQ). These
molecules are believed to be the basic building blocks of the
eumelanins, a class of bio-macromolecules with important
biological functions (including photoprotection) and with
potential for certain bioengineering applications. We have used
the $\Delta$SCF (difference of self consistent fields) method to
study the energy gap between the highest occupied molecular
orbital (HOMO) and the lowest unoccupied molecular orbital (LUMO),
$\Delta_\textrm{HL}$. We show that $\Delta_\textrm{HL}$ is similar
in IQ and SQ but approximately twice as large in HQ. This may have
important implications for our understanding of the observed broad
band optical absorption of the eumelanins. The possibility of
using this difference in $\Delta_\textrm{HL}$ to molecularly
engineer the electronic properties of eumelanins is discussed. We
calculate the infrared and Raman spectra of the three redox forms
from first principles. Each of the molecules have significantly
different infrared and Raman signatures, and so these spectra
could be used \textit{in situ} to non-destructively identify the
monomeric content of macromolecules. It is hoped that this may be
a helpful analytical tool in determining the structure of
eumelanin macromolecules and hence in helping to determine the
structure-property-function relationships that control the
behaviour of the eumelanins.
\end{abstract}

\maketitle

\section{Introduction}

The melanins are an important class of pigmentary macromolecule
found throughout the biosphere \cite{Prota}. Pheomelanin (a
cysteinyl-dopa derivative) and eumelanin (formed from
5-6-dihydroxyindolequinone and other indolequinones) are the
predominant forms in humans, and act as the primary
photoprotectant in our skin and eyes. Consistent with this role,
all melanins show broad band monotonic absorption in the UV and
visible in the range 1.5 to 5 eV \cite{Wolbarsht}. In contrast,
other biomolecules such as proteins and nucleic acids show only
well defined absorption peaks around 280 nm (4.5 eV) and little
absorption below that \cite{vanHolde}. They are also efficient
free radical scavengers and antioxidants \cite{Prota}. In direct
contradiction with these photoprotective properties, both
pheomelanin and eumelanin are implicated in the development of
melanoma skin cancer \cite{Hill_mel}. For this reason, the
photophysics, photochemistry and photobiology of melanins are
subjects of intense scientific interest.

Despite work over several decades (with respect to eumelanin in
particular), the more general structure-property-function
relationships that control the behaviour of these important
bio-macromolecules are still poorly understood \cite{Stark}. The
melanins are difficult molecules to study: they are chemically and
photochemically stable, and are virtually insoluble in most common
solvents. It is fairly well accepted that eumelanins are
macromolecules of the various redox forms of
5,6-dihydroxyindolequinone (DHI or HQ) and 5,6-dihydroxyindole
2-carboxylic acid (DHICA) \cite{Prota,Ito}. However, major
questions still remain concerning their basic structural unit
\cite{Zajac}. Two opposing schools of thought exist: i) that
eumelanins are composed of highly cross-linked extended
hetero-polymers based upon the Raper-Mason scheme \cite{Prota},
and ii) that eumelanins are actually composed of much smaller
oligomers condensed into 4 or 5 oligomer nano-aggregates
\cite{Clancy}. Clearly, this is a fundamental issue, and is the
starting point for the construction of consistent
structure-property-function relationships.

The answer to this question also has profound implications for our
understanding of the condensed phase properties of melanins. In
1960 Longuet-Higgins \cite{Longuet-Higgins}, in a landmark paper,
proposed that many of the physical properties of melanins could be
understood if they were semiconductors. This proposition was lent
further support by Pullman and Pullman \cite{Pullman} in 1964 who
applied molecular orbital theory in its simplest form (the
H\"uckel approximation) to 5,6-dihydroxyindole and one particular
dimer type. This theoretical work was followed in 1974 by the
experimental observations of McGinness, Corry and Proctor who
demonstrated that a pellet of melanin could be made to behave as
an amorphous electrical switch \cite{McGinness}. They postulated
that these materials may indeed be disordered organic
semiconductors consistent with the Longuet-Higgins theory, and the
newly developed models of amorphous inorganic semiconductors
\cite{Mott-Davis}. Several studies since have also claimed to show
that melanins in the condensed solid-state are semiconductors
\cite{Crippa,Jastrzebska}. However, it is by no means certain that
the conductivity reported in any of these experimental studies is
electronic in nature. A clear idea of the basic structural unit is
critical to developing a consistent model for condensed phase
charge transport in such disordered organic systems. It is also
important in the context of \lq\lq molecularly engineering"
melanins to have the ability to create or enhance functionality in
high technology applications such as bio-sensors and bio-mimetic
photovoltaics \cite{Meredith&Riesz03,JennyHonours}.

It is well known that disorder plays a crucial role in determining
the charge transport properties of amorphous inorganic
semiconductors. In addition to the usual impurity related
disorder, organic semiconductors (of which eumelanin may well be
an exotic example), present several different sources of disorder
associated with structural heterogeneity. For example, if one
considers either the hetero-polymer or oligomer models, there are
many different ways of constructing a eumelanin macromolecule
based upon the monomer sequences, cross-linking positions or
tautomer combinations. These possibilities have been studied
theoretically for DHI-based macromolecules using a number of
quantum chemical techniques. Notably, Galvao and Caldas
\cite{GC1,GC2,GC3} used the H\"uckel approximation to construct
model homo-polymers. Intriguingly, they found that many of the
important \lq\lq semiconductor related properties of eumelanin"
stabilise at a relatively small number of monomer units - five or
six in fact. Although they did not consider inter-chain effects,
and essentially treated chain terminations as end defects, their
work was the first indication that large, extended
hetero-polymeric structures are not required to explain the
physical properties of melanins. Bolivar-Marinez \etal
\cite{Bolivar-Marinez} and Bochenek and Gudowska-Nowak
\cite{Bochenek} have used the intermediate neglect of differential
overlap (INDO) and other semi-empirical methods to perform similar
calculations for oligomers, whilst other authors have utilised the
power of density functional theory to perform first principles
calculations \cite{Stark,Ilichev} on single monomers or dimers.

\begin{figure}
    \centering
    \epsfig{figure=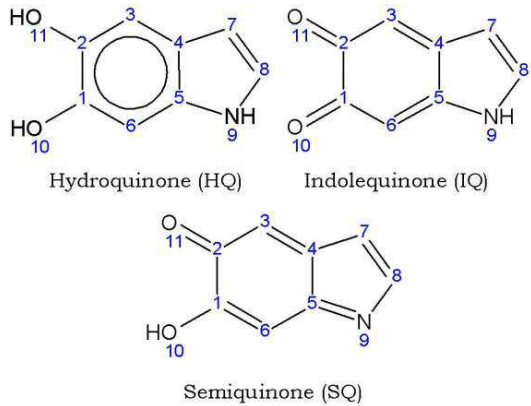, width=7cm, angle=0}
    \caption{Schematic representation of the 5,6-dihydroxyindolequinone (DHI) redox forms: hydroquinone (HQ)
    and the tautomers
    indolequinone (IQ) and semiquinone (SQ). The numbers correspond to those used in table \ref{tab:bond_lengths}.} \label{fig:struct}
\end{figure}

Il'ichev and Simon \cite{Ilichev} considered the hydroquinone,
(HQ) and indolequinone (IQ) forms (see figure \ref{fig:struct}),
and determined them to be the most stable and hence the most
likely forms of the monomers. However, several pulse radiolysis
studies
\cite{Lambert89,Lambert90,Al-Kazwini90,Al-Kazwini91,Al-Kazwini92a,Al-Kazwini92b}
have indicated the presence of the semiquinone (SQ - also shown in
figure \ref{fig:struct}), a tautomer of IQ, and so we also
consider SQ here.

Motivated by the remarkable optical absorption characteristics of
eumelanins, many of these authors have focused on simulating the
absorption spectra of HQ, IQ, SQ and related small oligomers
\cite{Ilichev,Bolivar-Marinez,Stark}. However, knowledge of the
electronic states around the highest occupied molecular orbital
(HOMO) and lowest unoccupied molecular orbital (LUMO) levels is
only part of the story. A great deal of information is also
contained within the vibrational and rotational structure of the
molecules. In particular, the IR and Raman spectra may be useful
analytical tools. This has proven to be the case in other
biomolecules \cite{vanHolde}. They can be used to confirm the
accuracy of the first principles predictions by direct comparison
with experimental evidence, and also to provide useful insight
into local environment phenomena such as solvent-solute
interactions. Furthermore, our IR and Raman spectra calculations
have the advantage of coming from first principles calculations
and not the semi-empirical methods often employed to calculate the
absorption spectra.

In this paper we present a full density functional theory (DFT)
analysis of the electronic, vibrational and rotational structure
of the indolequinone (IQ), the semiquinone (SQ) and the
hydroquinone (HQ). Our study is a considerable extension of
previously published work since we: i) present calculated IR and
Raman spectra for the first time, and ii) use the difference of
self consistent fields ($\Delta$SCF) approach to gain a more
accurate understanding of the LUMO level. We find that the
electronic properties of the molecules are highly dependent upon
the redox form. Hence, we can say that oligomers or
hetero-polymers consisting of one or more redox forms will have
large variations in the local chemical potential, or, in the
language of the tight binding model, strongly random site energies
and possibly also random hopping integrals. This is to say that
such hetero-structures will be highly disordered.

Our work is motivated by the desire to understand the implications
of the redox form of the basic structural unit of eumelanins on
the materials bulk properties. Such knowledge is a key starting
point in any attempts to understand charge transport in and the
optical properties of these disordered heterogeneous organic
conductors. It also has profound implications in unravelling the
mysteries of melanin biological functionality, and attempts to
molecularly engineer melanin-like molecules for technological
applications. These preliminary theoretical calculations are part
of an ongoing quantum chemical, experimental solid-state and
spectroscopic program aimed at gaining a more complete
understanding of melanin structure-property-function
relationships.

\section{Calculation details}

The chemical and electronic structures and the Raman and IR
spectra were found from first principles DFT calculations. We have
performed our calculations using the Naval Research Laboratory
Molecular Orbital Library (NRLMOL)
\cite{NRLMOL1,NRLMOL2,NRLMOL3,NRLMOL4,NRLMOL5,NRLMOL6,NRLMOL7}.
NRLMOL performs massively parallel electronic structure
calculation using gaussian orbital methods. In particular, for a
discussion of the calculation of the Raman and IR spectra see
Refs. \cite{NRLMOL5,NRLMOL6,NRLMOL7}. Throughout we have used the
Perdew, Burke and Ernzerhof (PBE) \cite{PBE} exchange correlation
functional, which is a generalised gradient approximation (GGA)
containing no parameters. 
For each molecule we have fully relaxed the geometry with no
symmetry constraints.

\section{Results and Discussion}\label{sect:results}

To benchmark our calculations we report the bond lengths and bond
angles of HQ, IQ and SQ found in our calculations in table
\ref{tab:bond_lengths}. Our results are in good agreement with
those calculated by Bolivar-Marinez \etal \cite{Bolivar-Marinez}
and Stark \etal \cite{Stark}. All three molecules are planar to
within numerical accuracy (c.f., \cite{Ilichev,Stark}).

We find, in agreement with Il'ichev and Simon \cite{Ilichev}, that
SQ is not the lowest energy tautomer of IQ (see table
\ref{tab:tautomers}). Indeed we find that at 300~K there should be
less than 0.1\% SQ. Although we should stress that this
calculation is for a single molecule \emph{in vacuo} it is, of
course, in direct contradiction to the evidence
\cite{Lambert89,Lambert90,Al-Kazwini90,Al-Kazwini91,Al-Kazwini92a,Al-Kazwini92b}
that SQ is one of the building blocks of eumelanin. However, since
SQ is widely thought to play an important role in eumelanin, we
also consider it here.

\begin{table}
\caption{\label{tab:tautomers} The total energy (in eV) of the two
lowest energy tautomers of indolequinone (IQ); semiquinone (SQ)
and 2-semiquinone (2-SQ) which is simply SQ with the hydrogen atom
bonded to O11 rather than O10 (c.f. figure \ref{fig:struct}). The
energies are quoted relative to the energy of IQ. In each case we
fully relaxed the geometries as shown in figures
\ref{fig:IQ_density} and \ref{fig:SQ_density}. Our results for HQ,
IQ and SQ are consistent with the calculations of Il'ichev and
Simon \cite{Ilichev}. The concentrations at 300~K are in direct
contradiction to the experimental evidence
\cite{Lambert89,Lambert90,Al-Kazwini90,Al-Kazwini91,Al-Kazwini92a,Al-Kazwini92b}
that SQ is one of the building blocks of eumelanin.}
\begin{ruledtabular}
\begin{tabular}{lccc}
Molecule & \begin{tabular}{c} Name in \\ Ref. \onlinecite{Ilichev}  \end{tabular}
& Energy (eV) & \begin{tabular}{c} Concentration \\ at $T=300$~K \end{tabular}\\
 \hline\vspace*{-9pt} \\
IQ & 3a & 0 & 87\% \\
2-SQ & 3d & 0.05 & 13\% \\
SQ & 3c & 0.19 & $<0.1\%$ \\
\end{tabular}
\end{ruledtabular}
\end{table}

\begin{figure}
    \centering
    \epsfig{figure=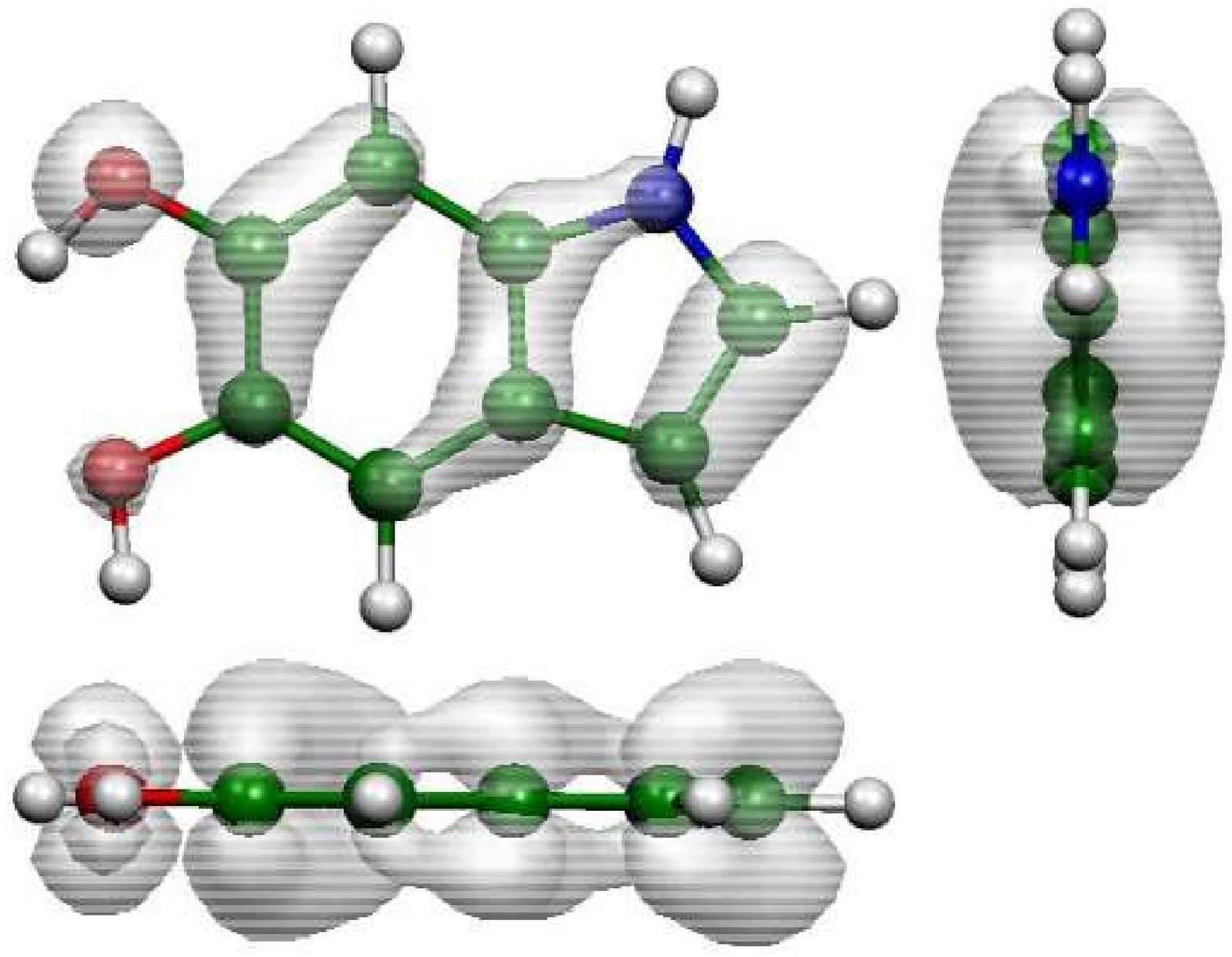, width=5cm, angle=0}
    \epsfig{figure=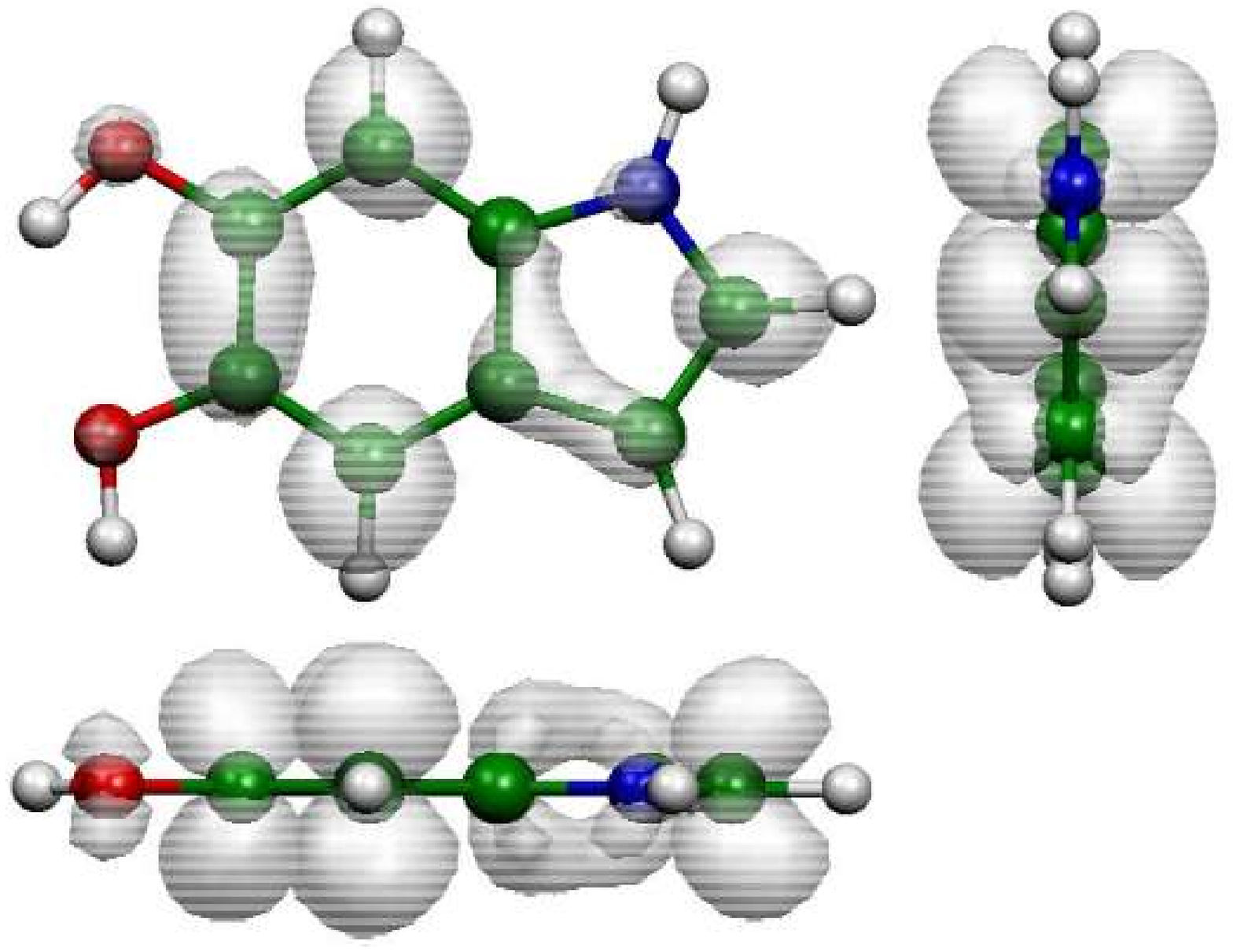, width=5cm, angle=0}
    \caption{The electron density in the highest occupied molecular
        orbital (HOMO) (top) and the lowest unoccupied molecular
        orbital LUMO (bottom) of hydroquinone (HQ). The atoms are
        colour coded as follows: carbon - green, nitrogen - blue,
        oxygen - red and hydrogen - white. The LUMO electron density is in good agreement
        with the semiempirical calculations of Bol\'iar-Marinez
        \textit{et al}. \cite{Bolivar-Marinez}, however there is some
        discrepancy in the HOMO electron density.}
    \label{fig:HQ_density}
\end{figure}

\begin{figure}
    \centering
    \epsfig{figure=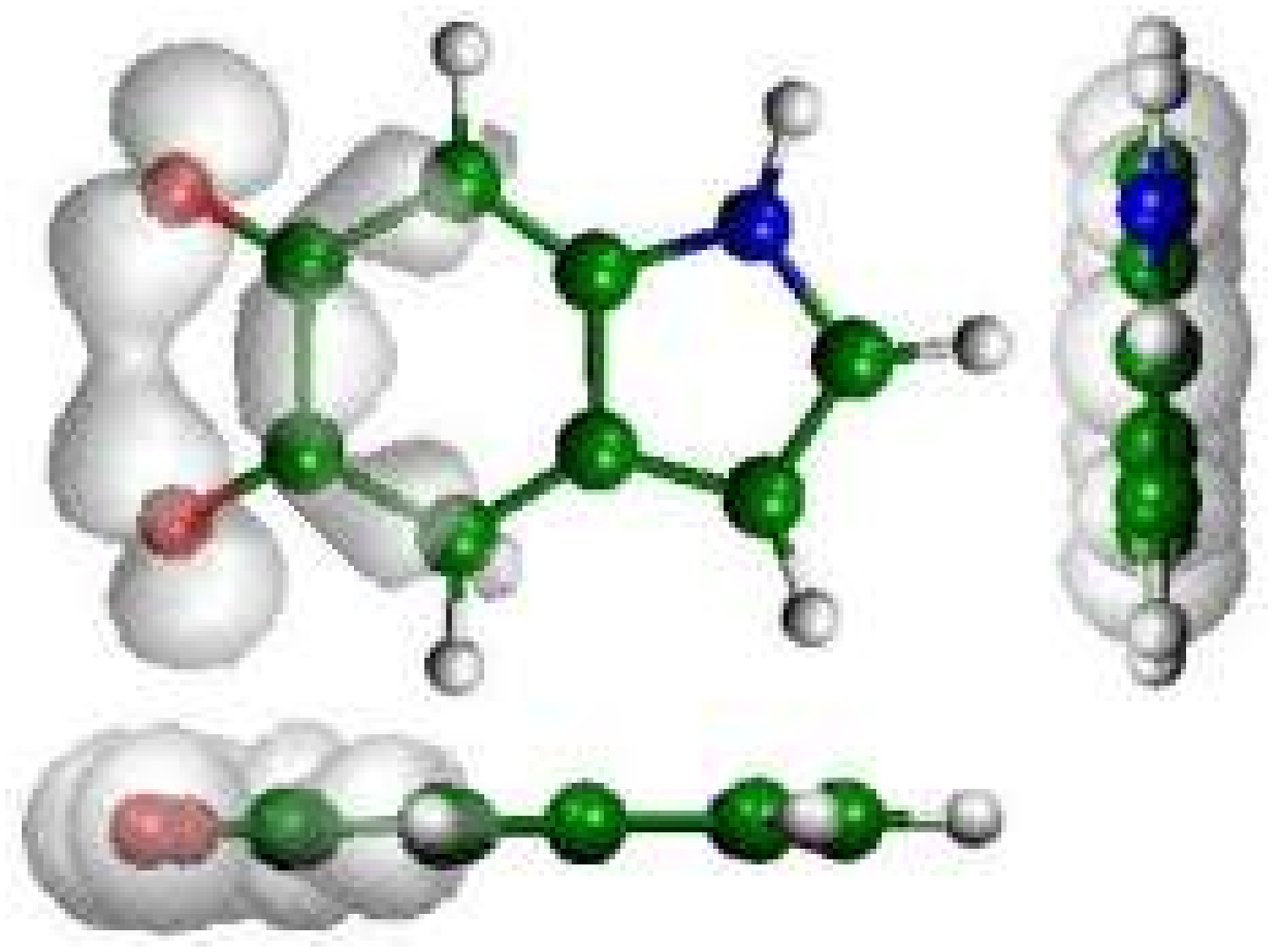, width=5cm, angle=0}
    \epsfig{figure=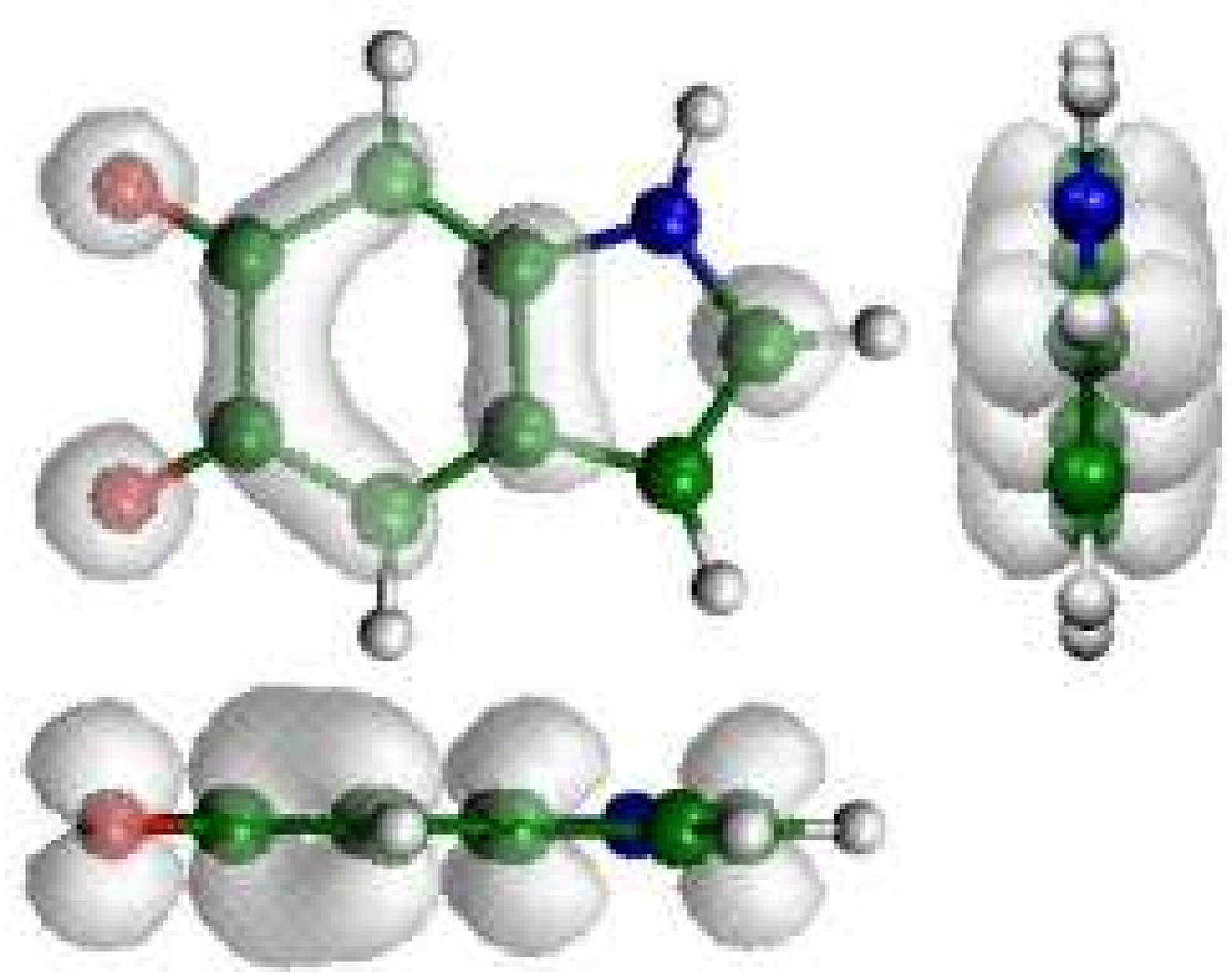, width=5cm, angle=0}
    \caption{The electron density in the highest occupied molecular
        orbital (HOMO) (top) and the lowest unoccupied molecular
        orbital LUMO (bottom) of indolequinone (IQ). The atoms are
        colour coded as follows: carbon - green, nitrogen - blue,
        oxygen - red and hydrogen - white. The LUMO electron density is in good agreement
        with the semiempirical calculations of Bol\'iar-Marinez
        \textit{et al}. \cite{Bolivar-Marinez}, however there is some
        discrepancy in the HOMO electron density.} \label{fig:IQ_density}
\end{figure}

\begin{figure}
    \centering
    \epsfig{figure=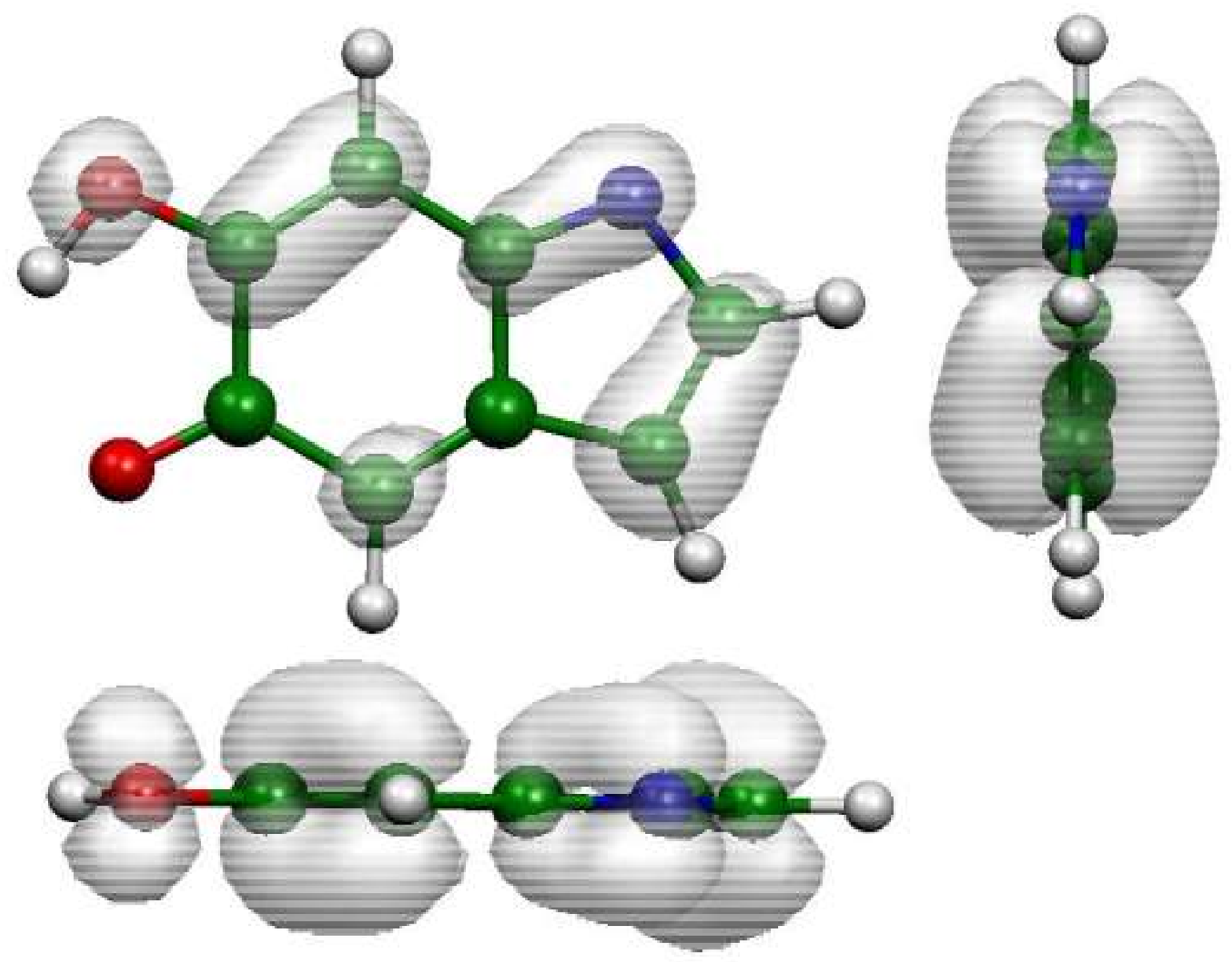, width=5cm, angle=0}
    \epsfig{figure=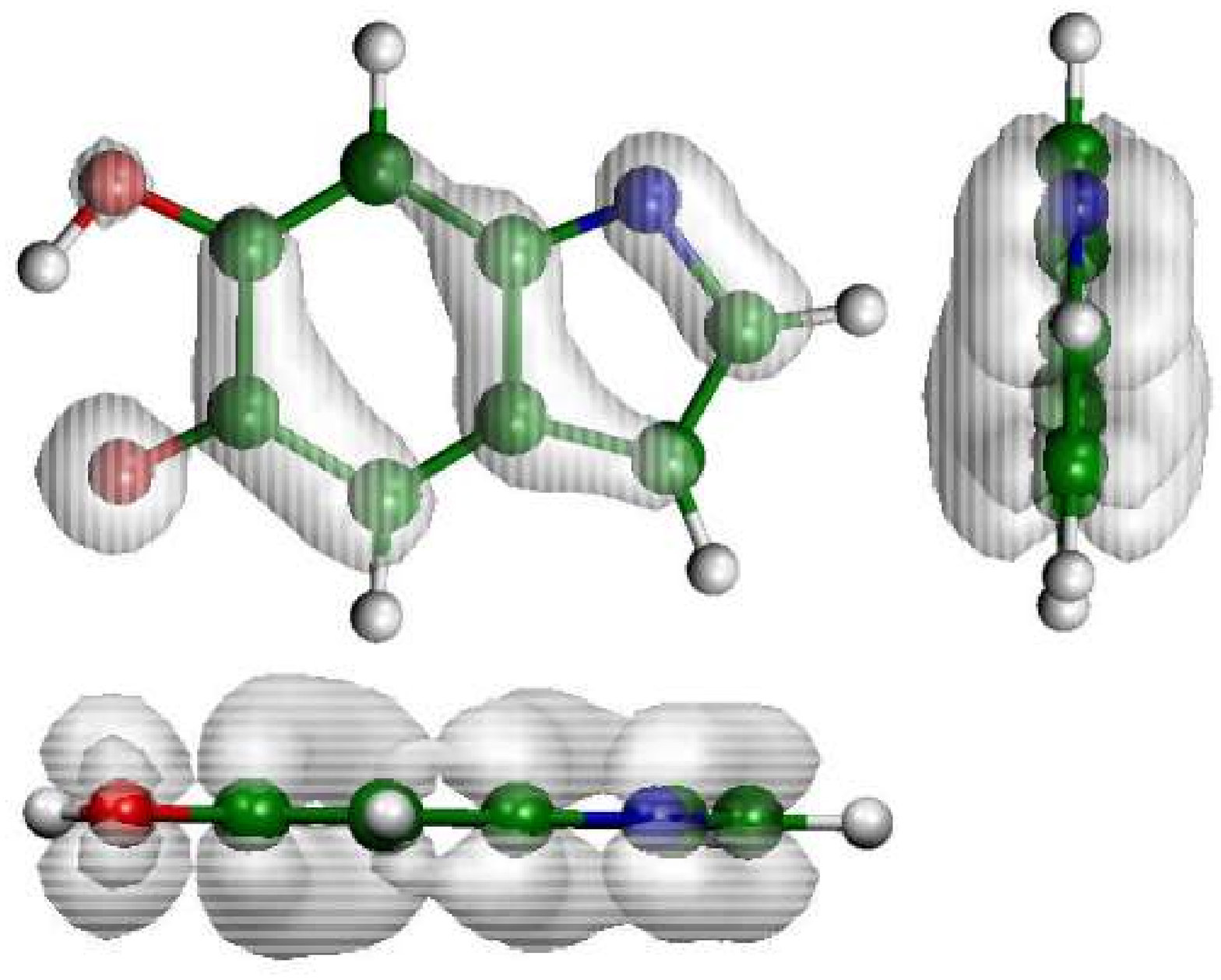, width=5cm, angle=0}
    \caption{The electron density in the highest occupied molecular
        orbital (HOMO) (top) and the lowest unoccupied molecular
        orbital LUMO (bottom) of semiquinone (SQ). The atoms are
        colour coded as follows: carbon - green, nitrogen - blue,
        oxygen - red and hydrogen - white. Both the HOMO and LUMO
        electron densities are in good agreement
        with the semiempirical calculations of Bol\'iar-Marinez
        \textit{et al}. \cite{Bolivar-Marinez}.} \label{fig:SQ_density}
\end{figure}

\begin{figure}
    \centering
    \epsfig{figure=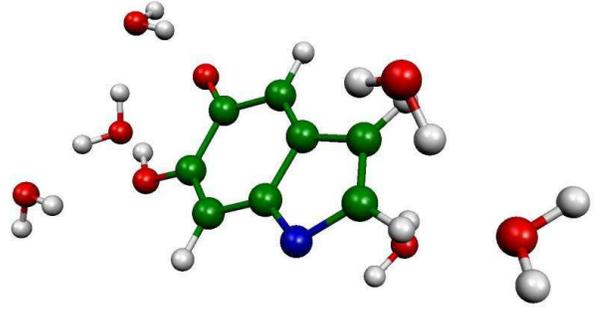, width=8cm, angle=0}
    \caption{The stable structure of semiquinone (SQ) and 6 H$_2$O molecules.
    Note the hydrogen bonded networks around the H-O-C-C=O group, which is dragged
    slightly out of the plan by the interactions with the solvent. The
    H-O-C bond angle has increased relative to that of the SQ molecule in vacuo.}
    \label{fig:SQ+6H2O}
\end{figure}

As well as the ground state geometries shown in figures
\ref{fig:HQ_density}, \ref{fig:IQ_density} and
\ref{fig:SQ_density} we also found stable geometries of HQ and SQ
with the H-O-C bond angle increased by almost 180$^\circ$. These
alternative structures have significantly higher energies than the
ground states. However, it is possible that the high energy
structures may be stabilised by hydrogen bonding in a polar
solvent. We therefore also considered the system
$\textrm{SQ}+6\textrm{H}_2\textrm{O}$. We found that in this
system the H-O-C bond is slightly larger than in SQ in vacuo (see
figure \ref{fig:SQ+6H2O}). We also found a significant hydrogen
bonded network around the H-O-C-C=O group which is dragged out of
the plane of the molecule by interactions with the water
molecules. This hydrogen bonding network is probably related to
the fact that eumelanins are only soluble in polar solvents.

\begin{table}
\caption{\label{tab:bond_lengths} The calculated bond lengths (in
\AA) and bond angles (in degrees) are found to be in good
agreement with previous work \cite{Bolivar-Marinez,Stark}. (Note
that there is no experimental data to compare with.) The atom
numbers correspond to those shown in figure \ref{fig:struct}.}
\begin{ruledtabular}
\begin{tabular}{lccc}
 & IQ & SQ  & HQ  \\
 \hline\vspace*{-9pt} \\
C1-C2 &  1.423  & 1.513 &  1.586 \\
C2-C3 &  1.388  & 1.486 &  1.458 \\
C3-C4 &  1.408  & 1.351 &  1.366 \\
C4-C5 &  1.421  & 1.488 &  1.483 \\
C5-C6 &  1.401  & 1.435 &  1.361 \\
C6-C1 &  1.392  & 1.362 &  1.457 \\
C4-C7 &  1.435  & 1.457 &  1.443 \\
C7-C8 &  1.376  & 1.358 &  1.365 \\
C8-N9 &  1.385  & 1.435 &  1.390 \\
N9-C5 &  1.382  & 1.310 &  1.381 \\
C1-O10 & 1.375  & 1.353 &  1.226 \\
C2-O11 & 1.376  & 1.226 &  1.226 \\
C1-C2-C3 &   120.69 & 117.88 &  118.32 \\
C2-C3-C4 &   119.89 & 119.61 & 119.72 \\
C3-C4-C5 &   118.45 & 120.78 & 121.03 \\
C4-C5-C6 &   122.20 & 121.32 & 124.60 \\
C5-C6-C1 &   118.18 & 118.79 & 118.34 \\
C6-C1-C2 &  120.59 & 121.62 & 117.98 \\
C3-C4-C7 &   134.47 & 134.88 & 132.73 \\
C6-C5-N9 &  130.61 & 127.05 & 129.38 \\
C4-C7-C8 &   107.08 & 105.05 & 107.23 \\
C7-C8-N9 &  109.33 & 113.75 & 110.86 \\
C8-N9-C5 &   109.32 & 105.23 & 109.65 \\
N9-C5-C4 &  107.19 & 111.63 & 106.02 \\
C5-C4-C7 &   107.08 & 104.34 & 106.24 \\
O10-C1-C6 &123.16 & 125.59 & 122.87 \\
O10-C1-C2 &  116.24 & 112.79 & 119.15 \\
O11-C2-C1 & 116.15 & 120.98 & 118.89 \\
O11-C2-C3 &  123.16 & 121.14 & 122.79
\end{tabular}
\end{ruledtabular}
\end{table}

The H\"uckel model calculations of Galvao and Caldas \cite{GC1}
indicate that the HOMO-LUMO gap ($\Delta_\textrm{HL}$) may be
significantly larger in HQ than in either IQ or SQ. This has
important implications in the context of understanding the basic
eumelanin structural unit. Fundamentally, a combination of HQ and
either IQ or SQ tautomers are present in an ensemble. Whether that
ensemble be oligomeric or polymeric it can be expected to lead to
an apparent broadening of the absorption profile. In this
situation, the key question is: how many monomers of HQ, IQ and SQ
are required to create the broad band UV and visible absorbance
over the range 1.5 to 5 eV that is macroscopically observable for
eumelanins? The answer to this question may well allow us to
assess the feasibility of the various structural models.
Furthermore, when one considers the possibility that molecularly
engineered forms of eumelanin may be useful as functional
materials in electronic devices and sensors
\cite{Meredith&Riesz03,JennyHonours}, a critical design parameter
is likely to be the \lq\lq semiconductor" gap. It has previously
been shown \cite{GC2} that, at least within the H\"uckel
approximation for homopolymers, the HOMO-LUMO gap of the monomer
is closely related to the HOMO-LUMO gap of the infinite polymer,
which, for the polymer model in the condensed solid state,
corresponds to the semiconducting gap provided there is only weak
coupling between polymers. Thus controlling the HOMO-LUMO gap of
macromolecules may provide a route to controlling the
semiconducting gap of eumelanins. This suggestion clearly needs
further investigation, but controlling the HQ to IQ or SQ ratio
may be a possible route to this form of property manipulation. An
important first step to investigating either of these proposals is
the calculation of the HOMO-LUMO gaps of the molecules by more
reliable methods than H\"uckel theory.

In table \ref{tab:HOMO-LUMO} we compare the HOMO-LUMO gap found in
our calculations with that found from the H\"uckel method
\cite{GC1}. For comparison with the H\"uckel calculations we need
to estimate the hopping integral, $\beta$, in the H\"uckel
Hamiltonian,
\begin{eqnarray}
\hat{\cal{H}} = \sum_{i\sigma}(\alpha+\alpha'\delta_{i9})\hat
c^\dagger_{i\sigma}\hat c_{i\sigma} - \beta\sum_{\langle
ij\rangle\sigma}\hat c^\dagger_{i\sigma}\hat c_{j\sigma},
\label{eqn:Huckel}
\end{eqnarray}
where $\alpha$ is the on-site energy, $\alpha'$ is the difference
in the on-site energy between a nitrogen and a carbon atom (we
take the labelling convention from figure \ref{fig:struct}
ensuring that the only nitrogen is in position $i=9$), $\hat
c^{(\dagger)}_{i\sigma}$ annihilates (creates) an electron with
spin $\sigma$ on site $i$ and $\langle ij\rangle$ indicates that
the sum is over nearest neighbours only. For hydrocarbons $\beta$
is typically of order 2.5~eV \cite{Lowe}. Using this value of
$\beta$ we find that the HOMO-LUMO gaps for HQ, IQ and SQ found
from the H\"uckel approximation are consistent with those found in
our DFT calculations using the PBE functional. However, because
DFT is a theory of the ground state, these calculations represent
the energy gap between Kohn--Sham eigenvalues and not the true
HOMO-LUMO gap of the molecules. This is known as the band gap
problem \cite{Jones&Gunnarsson}. Additionally, it is accepted that
the PBE functional can significantly underestimate the HOMO-LUMO
gap. Therefore we have also employed the $\Delta$SCF method
\cite{Jones&Gunnarsson} to calculate $\Delta_\textrm{HL}$. Our
results clearly reproduce the trends seen in the time dependent
density functional theory (TDDFT) calculations of Il'ichev and
Simon \cite{Ilichev} for IQ and SQ. The difference between the
$\Delta$SCF and TDDFT results for HQ probably indicates that the
true value of $\Delta_\textrm{HL}$ is intermediate to the values
predicted by the two different methods as it is extremely unlikely
that either method is incorrect by a large enough margin to allow
the other to be correct.

The calculated $\Delta$SCF HOMO-LUMO gap is also consistent with
previously published semiempirical ZINDO (Zerner's intermediate
neglect of differential overlap) calculations of the optical
absorption spectra \cite{Bolivar-Marinez,Stark}. In particular
Bol\'ivar-Marinez \etal noted that \lq\lq in the neutral state the
threshold for optical absorption is around 2.0~eV for the IQ and
SQ, while in the case of the HQ it is roughly 3.8~eV." Taking
these numbers as semiempirical estimates of the HOMO-LUMO gap they
are in excellent agreement with our $\Delta$SCF calculations (c.f.
table \ref{tab:HOMO-LUMO}).

It is also interesting to note that the change in the SQ
conformation observed in the calculations for the
$\textrm{SQ}+6\textrm{H}_2\textrm{O}$ system caused a 7\% decrease
in $\Delta_\textrm{HL}$ relative to SQ in vacuo. It is possible
that in a polar solvent a range of conformations exist which may
lead to a broadening of the optical absorption \cite{Myers}.

\begin{table}
\caption{\label{tab:HOMO-LUMO} The HOMO-LUMO (highest occupied
molecular orbital-lowest unoccupied molecular orbital) gap,
$\Delta_\textrm{HL}$, in eV. The HOMO-LUMO gap found from both the
$\Delta$SCF method and from a simple density functional
calculation (both calculations use the PBE functional \cite{PBE})
are compared with that found in the H\"uckel approximation (after
\cite{GC1}, taking $\beta=2.5$~eV which is the correct order of
magnitude \cite{Lowe}, c.f., (\ref{eqn:Huckel})). It can be seen
that the HOMO-LUMO gap of both the indolequinone (IQ) and the
semiquinone (SQ) is significantly underestimated by both the
H\"uckel approximation and the simple interpretation of the
Kohn-Sham eigenvalues. However, these methods find a HOMO-LUMO gap
for the hydroquinone (HQ) that is more consistent with the, more
reliable, $\Delta$SCF method. The results of the $\Delta$SCF
method indicate that the HOMO-LUMO gap is approximately the same
in IQ and SQ but about a factor of two larger in HQ. These results
correctly reproduce the trends seen in the more accurate, more
more computationally expensive time dependent density functional
theory (TDDFT) calculations of Il'ichev and Simon \cite{Ilichev}.
This is consistent with the threshold for optical absorption found
from semi-empirical methods \cite{Bolivar-Marinez,Stark}. The
large difference in the HOMO-LUMO gap between HQ and the other
molecules may be important for \lq\lq molecular engineering"
applications and for explaining the broad band optical absorption
of eumelanin. In particular, if the model of the eumelanins as
amorphous semiconductors is correct it may allow the chemical
tuning of the semiconducting band gap (c.f. Ref. \cite{GC2}).}
\begin{ruledtabular}
\begin{tabular}{lccc}
 & IQ & SQ  & HQ  \\
 \hline\vspace*{-9pt} \\
TDDFT (PBE/B3LYP) \cite{Ilichev}
& 1.82 & 1.50 & 4.53 \\
TDDFT (B3LYP) \cite{Ilichev}
& 1.79 & 1.43 & 4.30 \\
 NRLMOL ($\Delta$SCF/PBE)
&  2.02 & 1.12 & 3.61 \\
 NRLMOL (PBE)
&  1.07 & 0.80 & 3.48 \\
H\"uckel \cite{GC1}
& 1.3 & 0.84 & 3.4 \\
\end{tabular}
\end{ruledtabular}
\end{table}

In terms of H\"uckel theory, the difference between
$\Delta_\textrm{HL}$ for HQ and IQ/SQ can be understood in terms
of the level of delocalisation of the HOMOs and LUMOs. It should
be noted that the aromatic ring in HQ contains a resonating
valence bond, whereas this feature is absent in IQ and SQ. This
can be contrasted with the H\"uckel model result for benzene which
predicts $\Delta_\textrm{HL}=2\beta$ for the neutral molecule but
if one double bond in the ring is replaced by a single bond, as
in, for example, the reduced form C$_6$H$_8$, then
$\Delta_\textrm{HL}=0$. On this basis one might well expect the
HOMO-LUMO gap of HQ to be significantly larger than those of IQ
and SQ. However, detailed plots of the electron densities in the
HOMO and LUMO of the three molecules (figures
\ref{fig:HQ_density}, \ref{fig:IQ_density} and
\ref{fig:SQ_density}) indicate that the real situation is somewhat
more complicated. The electron densities are significantly more
localised in the HOMOs of IQ and SQ than the electron density of
the HOMO of HQ is. In particular, the electron densities in the
HOMOs and LUMOs of IQ and SQ are not as similar as one might
expect from figure \ref{fig:struct}. However, even in the
$\Delta$SCF calculation, the HOMO-LUMO gap is a factor of two
larger in HQ than it is in SQ and IQ.

\begin{table}
\caption{\label{tab:HQ} The predicted Phonon (IR) and Raman
spectra for hydroquinone (HQ).}
\begin{tabular}{cccc}
\\\hline \hline  \vspace*{-6pt}  \\
 \begin{tabular}{c} Frequency  \\ (cm$^{-1}$) \end{tabular}&
 \begin{tabular}{c} IR intensity \\ (D$^2$/amu\AA$^2$) \end{tabular}&
 \begin{tabular}{c} Isotropic Raman \\ scattering \\ activity \\
  (\AA$^4$/amu)\end{tabular}&
 \begin{tabular}{c} Total Raman \\ scattering \\ activity \\
  (\AA$^4$/amu) \end{tabular}  \vspace*{2pt}   \\
 \hline \vspace*{-7pt} \\
137.4   &   0.008   &   0.000   &   0.668   \\
144.0   &   0.162   &   0.000   &   0.605   \\
289.5   &   0.357   &   0.000   &   0.729   \\
300.7   &   0.389   &   0.022   &   0.908   \\
302.9   &   1.848   &   0.000   &   2.430   \\
311.7   &   0.976   &   0.000   &   1.205   \\
314.0   &   0.019   &   0.023   &   0.533   \\
344.3   &   0.932   &   0.000   &   0.040   \\
366.5   &   1.360   &   0.000   &   0.993   \\
416.0   &   0.145   &   0.000   &   0.990   \\
428.2   &   0.120   &   0.013   &   3.523   \\
463.2   &   0.032   &   9.290   &   14.050  \\
589.6   &   0.040   &   0.000   &   0.380   \\
608.8   &   0.561   &   0.000   &   1.277   \\
665.2   &   0.581   &   0.000   &   0.520   \\
677.4   &   0.376   &   0.000   &   0.626   \\
706.8   &   0.689   &   0.000   &   0.811   \\
735.0   &   0.011   &   31.978  &   36.825  \\
753.5   &   0.036   &   0.178   &   1.655   \\
762.5   &   0.250   &   0.000   &   0.181   \\
792.1   &   0.506   &   0.000   &   0.947   \\
806.4   &   0.436   &   0.000   &   0.776   \\
843.0   &   0.639   &   0.477   &   0.636   \\
890.6   &   0.125   &   2.517   &   4.183   \\
1055.2  &   0.291   &   21.980  &   28.451  \\
1069.8  &   0.129   &   0.213   &   2.528   \\
1104.2  &   2.271   &   1.698   &   2.285   \\
1145.8  &   0.635   &   0.034   &   9.197   \\
1159.9  &   8.026   &   0.708   &   1.406   \\
1175.8  &   2.970   &   0.000   &   3.764   \\
1217.1  &   0.303   &   1.202   &   2.299   \\
1265.2  &   0.391   &   5.638   &   8.478   \\
1328.6  &   0.909   &   26.707  &   45.852  \\
1334.9  &   1.440   &   3.164   &   4.150   \\
1360.6  &   0.213   &   68.155  &   122.963 \\
1425.4  &   0.135   &   1.951   &   12.385  \\
1462.3  &   2.384   &   3.954   &   38.040  \\
1489.8  &   0.324   &   0.016   &   0.614   \\
1511.3  &   1.122   &   64.105  &   92.156  \\
1593.6  &   0.436   &   0.005   &   1.408   \\
1628.4  &   0.366   &   3.559   &   21.170  \\
3086.8  &   0.425   &   30.403  &   58.297  \\
3090.6  &   0.257   &   90.643  &   170.535 \\
3170.7  &   0.047   &   2.066   &   86.139  \\
3189.4  &   0.033   &   128.345 &   185.587 \\
3598.1  &   1.359   &   78.683  &   135.123 \\
3709.9  &   1.153   &   74.758  &   126.550 \\
3716.9  &   1.182   &   98.150  &   155.123 \\
\hline\hline
\end{tabular}
\end{table}

\begin{table}
\caption{\label{tab:IQ} The predicted Phonon (IR) and Raman
spectra for indolequinone (IQ).}
\begin{ruledtabular}
\begin{tabular}{cccc}
 \begin{tabular}{c} Frequency  \\ (cm$^{-1}$) \end{tabular}&
 \begin{tabular}{c} IR intensity \\ (D$^2$/amu\AA$^2$) \end{tabular}&
 \begin{tabular}{c} Isotropic Raman \\ scattering \\ activity \\
  (\AA$^4$/amu)\end{tabular}&
 \begin{tabular}{c} Total Raman \\ scattering \\ activity \\
  (\AA$^4$/amu) \end{tabular}  \\
 \hline
55.0    &   0.004   &   0.000   &   0.612   \\
122.2   &   0.000   &   0.000   &   0.580   \\
197.4   &   0.292   &   0.000   &   1.234   \\
289.1   &   1.546   &   0.000   &   0.095   \\
329.2   &   0.026   &   0.161   &   1.736   \\
336.2   &   0.045   &   0.963   &   1.476   \\
357.9   &   0.152   &   0.000   &   0.251   \\
396.5   &   0.178   &   1.587   &   19.222  \\
445.9   &   0.002   &   0.000   &   0.249   \\
448.8   &   0.007   &   7.783   &   14.949  \\
568.5   &   0.143   &   0.000   &   0.531   \\
614.5   &   0.008   &   15.800  &   20.336  \\
643.6   &   0.515   &   0.994   &   14.159  \\
692.8   &   0.409   &   0.000   &   0.493   \\
735.5   &   0.175   &   0.000   &   0.209   \\
743.0   &   0.246   &   0.000   &   0.332   \\
746.1   &   0.069   &   1.722   &   14.246  \\
806.9   &   1.552   &   0.000   &   1.162   \\
814.5   &   0.049   &   4.421   &   8.129   \\
850.4   &   0.118   &   0.000   &   0.640   \\
850.7   &   0.260   &   0.001   &   1.761   \\
869.6   &   0.012   &   0.000   &   1.615   \\
1047.0  &   1.402   &   1.668   &   17.404  \\
1072.1  &   0.549   &   25.395  &   72.312  \\
1104.1  &   0.006   &   10.970  &   38.889  \\
1133.4  &   0.339   &   0.008   &   6.465   \\
1184.8  &   0.119   &   2.879   &   6.366   \\
1208.0  &   1.243   &   23.620  &   33.118  \\
1218.5  &   1.682   &   0.037   &   0.926   \\
1324.5  &   0.322   &   19.383  &   44.473  \\
1343.5  &   4.097   &   0.757   &   25.939  \\
1405.0  &   0.710   &   0.017   &   5.047   \\
1533.9  &   1.998   &   125.526 &   209.070 \\
1575.3  &   4.151   &   148.528 &   208.833 \\
1623.9  &   2.864   &   8.682   &   131.748 \\
1640.5  &   4.173   &   25.925  &   103.570 \\
1667.0  &   0.874   &   0.699   &   17.584  \\
3109.9  &   0.078   &   72.246  &   127.495 \\
3121.8  &   0.040   &   90.847  &   166.066 \\
3177.1  &   0.017   &   4.399   &   100.595 \\
3195.5  &   0.003   &   133.664 &   192.718 \\
3594.6  &   1.840   &   95.806  &   150.728 \\
\end{tabular}
\end{ruledtabular}
\end{table}

\begin{table}
\caption{\label{tab:SQ} The predicted Phonon (IR) and Raman
spectra for semiquinone (SQ).}
\begin{ruledtabular}
\begin{tabular}{cccc}
 \begin{tabular}{c} Frequency  \\ (cm$^{-1}$) \end{tabular}&
 \begin{tabular}{c} IR intensity \\ (D$^2$/amu\AA$^2$) \end{tabular}&
 \begin{tabular}{c} Isotropic Raman \\ scattering \\ activity \\
  (\AA$^4$/amu)\end{tabular}&
 \begin{tabular}{c} Total Raman \\ scattering \\ activity \\
  (\AA$^4$/amu) \end{tabular}  \\
 \hline
89.5    &   0.161   &   0.000   &   0.573   \\
136.3   &   0.004   &   0.000   &   0.242   \\
221.8   &   0.001   &   0.000   &   0.318   \\
307.7   &   0.164   &   1.094   &   1.619   \\
348.8   &   0.034   &   0.023   &   0.434   \\
352.9   &   0.025   &   0.000   &   0.121   \\
383.3   &   0.461   &   0.000   &   0.080   \\
402.1   &   0.041   &   0.057   &   0.561   \\
447.6   &   0.069   &   11.923  &   25.667  \\
458.3   &   2.024   &   0.000   &   1.902   \\
516.1   &   0.000   &   0.000   &   1.370   \\
628.7   &   0.770   &   2.541   &   5.507   \\
656.8   &   0.077   &   15.330  &   22.320  \\
686.7   &   0.337   &   0.000   &   0.263   \\
748.5   &   0.034   &   0.000   &   0.697   \\
753.8   &   0.051   &   3.268   &   4.895   \\
756.0   &   0.322   &   0.000   &   0.476   \\
813.9   &   0.045   &   0.071   &   1.579   \\
820.4   &   0.121   &   2.587   &   8.251   \\
833.0   &   0.422   &   0.000   &   1.924   \\
885.4   &   0.849   &   0.000   &   0.653   \\
902.3   &   0.053   &   0.000   &   1.303   \\
983.0   &   0.542   &   5.903   &   87.829  \\
1075.3  &   1.431   &   12.652  &   24.493  \\
1087.2  &   2.371   &   13.610  &   22.939  \\
1117.7  &   1.632   &   0.159   &   9.469   \\
1173.4  &   0.801   &   0.135   &   2.521   \\
1199.1  &   2.031   &   10.995  &   15.634  \\
1233.3  &   1.360   &   0.819   &   4.704   \\
1294.0  &   0.974   &   13.021  &   66.258  \\
1356.8  &   0.767   &   2.578   &   8.707   \\
1384.4  &   0.077   &   24.950  &   39.103  \\
1499.3  &   0.607   &   248.871 &   433.636 \\
1537.7  &   0.176   &   18.043  &   69.559  \\
1585.3  &   4.588   &   70.696  &   147.517 \\
1648.9  &   0.164   &   136.594 &   232.764 \\
1650.3  &   5.288   &   26.421  &   93.212  \\
3088.1  &   0.261   &   76.797  &   140.364 \\
3121.5  &   0.026   &   60.571  &   122.056 \\
3153.0  &   0.214   &   32.031  &   134.793 \\
3181.5  &   0.076   &   123.921 &   185.037 \\
3680.9  &   2.011   &   108.125 &   198.914 \\
\end{tabular}
\end{ruledtabular}
\end{table}

In the absence of knowledge of the chemical structure of the
eumelanins it is difficult to  predict how a change in the
HOMO-LUMO gap of the monomer will effect the electronic structure.
However it is likely that changing $\Delta_\textrm{HL}$ by a
factor of two will have a dramatic effect on the electronic
structure of the melanin macromolecule. For example, it is
reasonable to expect that the HOMO-LUMO gap of a macromolecule
will be related to $\Delta_\textrm{HL}$ (c.f. Ref. \cite{GC2}).
Thus we expect that the HOMO-LUMO gap for a macromolecule of HQ
will be significantly larger than that of a macromolecule of IQ or
SQ. For a macromolecule containing both HQ and, say, IQ this
change in the electronic structure will act as a source of
disorder and thus have a dramatic effect on the electronic
transport properties.

It is also difficult, without detailed knowledge of the chemical
structure of the eumelanins, to predict exactly how the variation
of $\Delta_\textrm{HL}$ affects the optical absorption. However,
it seems possible that the range of $\Delta_\textrm{HL}$ between
different molecules may produce the continuum of HOMO-LUMO gaps in
eumelanin macromolecules and thus play an important role in
explaining the observed the broad band optical absorption.


It is therefore important to be able to identify the monomeric
content of a sample of eumelanin, ideally this should be done
\textit{in situ} and non-destructively. Motivated by this fact we
have calculated both the Raman and infrared (IR) spectra of IQ, SQ
and HQ, (see tables \ref{tab:HQ}, \ref{tab:IQ} and \ref{tab:SQ}).
This has several uses. Firstly, the prediction allows for
experimental testing of the accuracy of our calculations.
Secondly, it can be seen that there are notable differences in
both the Raman and IR spectra of the three monomers (see, for
example, figures \ref{fig:IRzoom} and \ref{fig:RamanZoom}). The
Raman and IR spectra could therefore be used for the \textit{in
situ}, non-destructive identification of the monomeric content of
macromolecules. This could be particularly valuable in the
engineering of devices from eumelanins, as, by varying the ratio
of indole-quinones, it may be possible to achieve control of the
band-gap of the material. This may also be a useful analytical
tool for the analysis of the structure of both synthetic and
naturally occurring melanin, and thus be helpful in determining
the structure-property-function relationships that control their
behaviour.

Comparison of our calculated IR and Raman spectra with standard
tables for organic molecules \cite{Nakanishi&Solomon,Schrader}
shows that the calculated phonon spectra are in broad agreement
with known IR absorption band values (see table \ref{tab:IR}). It
can also be seen from table \ref{tab:IR} that the differences in
the spectra, specifically which vibrations are absent for a given
molecule can be understood in terms of their structure, c.f.
figure \ref{fig:struct}, as expected.

\begin{table*}
\caption{\label{tab:IR} Broad assignments of the strongest bands
in the simulated phonon (IR) spectra of HQ, SQ and IQ (FIG.6). The
band assignments were made according to standard tables
\cite{Nakanishi&Solomon,Schrader}.}
\begin{ruledtabular}
\begin{tabular}{lcccc}
Band (cm$^{-1}$) & Group/Vibration & HQ & SQ & IQ \\
 \hline\vspace*{-9pt} \\
$\sim$1100 & $\nu$ (C-OH) strong & present & present & absent \\
$\sim$1340 & $\nu$ (Aromatic C-N) strong & present & absent & present \\
$\sim$1450-1590 & $\nu$ (Aromatic C-C) strong & present & present & present \\
$\sim$1620-1650 & $\nu$ (quinone C=O) strong & absent & present & present \\
$\sim$3590-3600 & $\nu$ (N-H) medium  & present & absent & present \\
$\sim$3700 & $\nu$ (O-H) strong & present & present & absent
\end{tabular}
\end{ruledtabular}
\end{table*}

It is important to stress that our calculated IR and Raman spectra
are for gaseous phase monomers. Thus, the comparison of such
results with experiment may be complicated by several factors.
Firstly the formation of macromolecules may result in a
significant change in the spectra. However, it is known that in
other biomolecules the effect of polymerisation only slightly
broadens the IR and Raman spectra, for example the spectra of
simple amino acids is only slightly broadened in proteins
\cite{Xie}. Secondly solvent effects may largely \lq wash out' the
individual features of the spectra \cite{Reichardt,Myers}. It is
clear from our study of the $\textrm{SQ}+6\textrm{H}_2\textrm{O}$
system that there will be significant hydrogen bonding in the
solvent, hydrogen bonding can shift the phonon frequencies
\cite{vanHolde}. Therefore it remains to be seen if,
experimentally, the resolution is good enough to resolve the
differences in the spectra of the monomers. However, an obvious
way to avoid solvent effects is to conduct the experiments in the
solid state.

\section{Conclusions}

We have carried out first principles density functional
calculations for the hydroquinone (HQ), the indolequinone (IQ) and
the semiquinone (SQ). The calculated gaseous phase structure is in
good agreement with previous calculations. We have used the
$\Delta$SCF method to study the HOMO-LUMO gap. Our results are
consistent with the threshold for optical absorption found by
semiempirical methods. Specifically, we found that the HOMO-LUMO
gap is similar in IQ and SQ but approximately twice as large in
HQ. The possibility of using this difference in the HOMO-LUMO gap
to engineer the electronic properties of eumelanins at the
molecular level has been discussed. We have also suggested that
the difference in the HOMO-LUMO gap of the different monomers
could lead to a large range of HOMO-LUMO gaps in eumelanin
macromolecules and thus be related to the observed broad band
optical absorption.

As the structure of macromolecules of these monomers is not known
we have also calculated the IR and Raman spectra of the three
monomers from first principles. A comparison of these results with
experiment would represent a stringent test of the density
functional calculations as these calculations do not rely on
additional semiempirical information. It was shown that the IR and
Raman spectra have potential as a analytical tool for these
materials. Each of the monomers have significantly different
spectral signatures, therefore the IR or Raman spectra could be
used \textit{in situ} to non-destructively investigate the
monomeric content of macromolecules. It is hoped that this may be
helpful in determining the structure of both natural and synthetic
eumelanins, and hence aid our attempts to understand their
biological functionality.

\section*{Acknowledgements}

We would like to thank John Dobson, Joel Gilmore, Peter Innis,
Urban Lundin, Jeffrey Reimers, Jennifer Riesz Cathy Stampfl and
Linh Tran for helpful discussions. The work at the University of
Queensland was supported by the Australian Research Council. BJP
was supported in part by US Navy Grant N00014-03-1-4115. KB was
supported in part by US Navy Grant N00014-03-1-4116. TB was
supported in part by US Navy Grant N00014-02-1-1046. MRP was
supported in part by ONR and the DoD HPC CHSSI initiative. BJP,
RHM and MRP appreciate the hospitality of the Kavli Institute of
Theoretical Physics, where this work was initiated. The KITP is
supported by NSF through grant PHY99-07949.

\bibliographystyle{apsrev}
\bibliography{melanin}

\begin{widetext}


\begin{figure}[p]
    \centering
    \epsfig{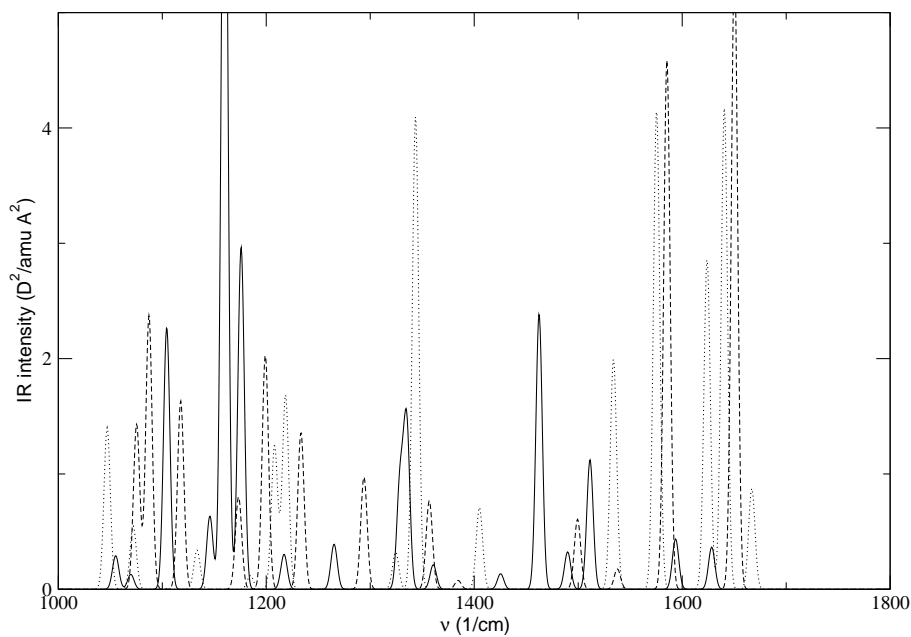}
    \caption{Detail of the phonon (IR) spectra of hydroquinone (HQ)
    (solid line), semiquinone (SQ) (dashed line) and indolequinone
    (IQ) (dotted line). Clear differences can be seen between the
    spectra of the three monomers. For example, there are strong
    features around 1100cm$^{-1}$ in both HQ and SQ but not in IQ. On
    the other hand, there are strong bands between $\sim$1620 and
    1650cm$^{-1}$
    in SQ and IQ which are absent in HQ (for broad assignments see table \ref{tab:IR}).
    This confirms that the IR spectra could be useful as in situ,
    non-destructive probes for identifying the monomeric content
    of eumelanins. To allow us to plot the spectra and to aid
    comparison with experiment we have broadened each peak by 6cm$^{-1}$,
    a value that is comparable to that seen experimentally for other
    biomolecules in solution \cite{vanHolde}.} \label{fig:IRzoom}
\end{figure}


\begin{figure}[p]
    \centering
    \epsfig{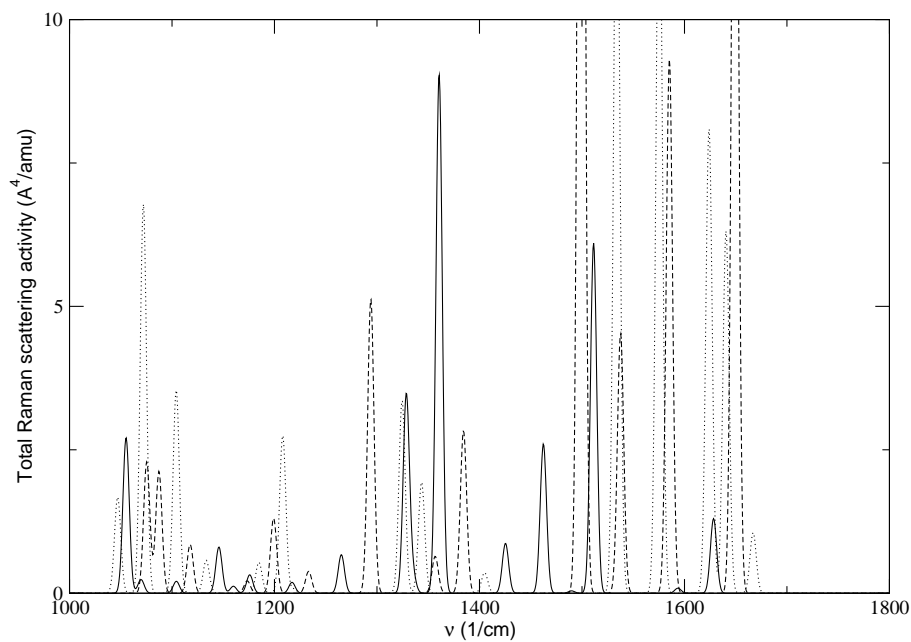}
    \caption{Detail of the Raman spectra of hydroquinone (HQ)
    (solid line), semiquinone (SQ) (dashed line) and indolequinone
    (IQ) (dotted line). Once again, there are clear differences
    between the spectra of the three monomers confirming that
    Raman spectra could be useful as in situ, non-destructive
    probes for identifying the monomeric content of eumelanins.
    To allow us to plot the spectra and to aid comparison with
    experiment we have broadened each peak by 6cm$^{-1}$,
    a value that is comparable to that seen experimentally for other
    biomolecules in solution \cite{vanHolde}.} \label{fig:RamanZoom}
\end{figure}

\end{widetext}
\end{document}